\documentclass[prb,twocolumn]{revtex4-1}
\usepackage{graphicx}
\usepackage{hyperref}
\usepackage{amssymb}
\usepackage{dcolumn}
\usepackage{float}
\usepackage{bm}
\usepackage{multirow}
\usepackage{booktabs}
\usepackage[T1]{fontenc}
\usepackage[utf8]{inputenc}
\usepackage{lmodern}
\usepackage{color}
\usepackage{makecell}
\usepackage{tikz}
\usepackage{pgfplots}
\usepackage{mathtools}
\usepackage{amsmath} 

\newcommand{\mathleft}{\@fleqntrue\@mathmargin0pt}

\newcolumntype{M}[1]{>{\centering\arraybackslash}m{#1}}

\renewcommand{\i}{\mathrm{i}}

\newcommand{\tens}{%
	\mathbin{\mathop{\otimes}}%
}
\DeclareMathAlphabet{\bi}{OML}{cmm}{b}{it}

\def\be{\begin{equation}}
\def\ee{\end{equation}}
\def\bearr{\begin{eqnarray}}
\def\eearr{\end{eqnarray}}

\begin{document}
	
\title{Magnetotransport properties of the Quantum Spin Hall and Quantum Hall states in an inverted HgTe/CdTe and InAs/GaSb quantum wells
}
\bigskip
\author{Alestin Mawrie}
\normalsize
\affiliation{Department of Physics, Indian Institute of Technology Indore, Simrol, Indore-453552, India}
\date{\today}
\begin{abstract}
	The quantum spin Hall (QSH) states discovered in an inverted band of InAs/GaSb and HgTe/CdTe quantum wells categorize them among the very superior candidates for topological insulators. In the presence of a magnetic field, these QSH states persist up to a magnetic field equal to the critical field, beyond which the edge states would consist of normal quantum Hall (QH) states. We provide the expression of this critical field which is found consistent with some previous literature. The critical field partitioned the spectrum into two types of quantum states, \textit{viz}. , the Quantum spin Hall (QSH) and Quantum Hall (QH) states. We present a theoretical study of the magnetotransport properties based on the Bernevig-Hughes-Zhang Hamiltonian that describes these QSH states. Our results of the Hall conductivity show the different responses at these two different topological regions. Around the low Fermi energy level, the system has a high Hall conductivity in the QH region, while the same is less dominant in the QSH region. Our results of the Hall conductivity thus help differentiate the type topological phase of the given quantum well.
\end{abstract}

\email{amawrie@iiti.ac.in}
\pacs{78.67.-n, 72.20.-i, 71.70.Ej}

\maketitle
\section{Introduction}
A topological insulator is a material that behaves as an insulator in the bulk, but has conducting states at their edges or surfaces, meaning that electrons conduct dissipationless along the edges or surfaces of the material\cite{topo,topo2}. These surface/edge electronic states have spin-orientations following the direction of the electron's momentum, thus protected by the time-reversal (TR) symmetry\cite{TR}. Quantum spin Hall (QSH) states in HgTe/CdTe \cite{ref1,ref2,ref3,ref4} and InAs/GaSb \cite{InAs1,InAs2,ref2,InAs4} quantum wells arising from the band inversion (where the top of the valence band is at an energy level higher than the bottom of the conduction band) are a few of such candidates where topological properties are inevitable. The band inversion in these systems is tunable to achieve a topological insulating behavior\cite{tune,tune2}. 
The Fermi level in such quantum wells are accompanied by a completely inverted band structure\cite{bhz,Hamil2}.
Also, the QSH states in these materials are characterized by a topological $Z_2$ invariant that predicts their topological phase, (whether the system hosts the trivial insulating states or the QSH time-reversal symmetry protected edge states).

\begin{figure}[t]
	\includegraphics[width=42.5mm,height=35.5mm]{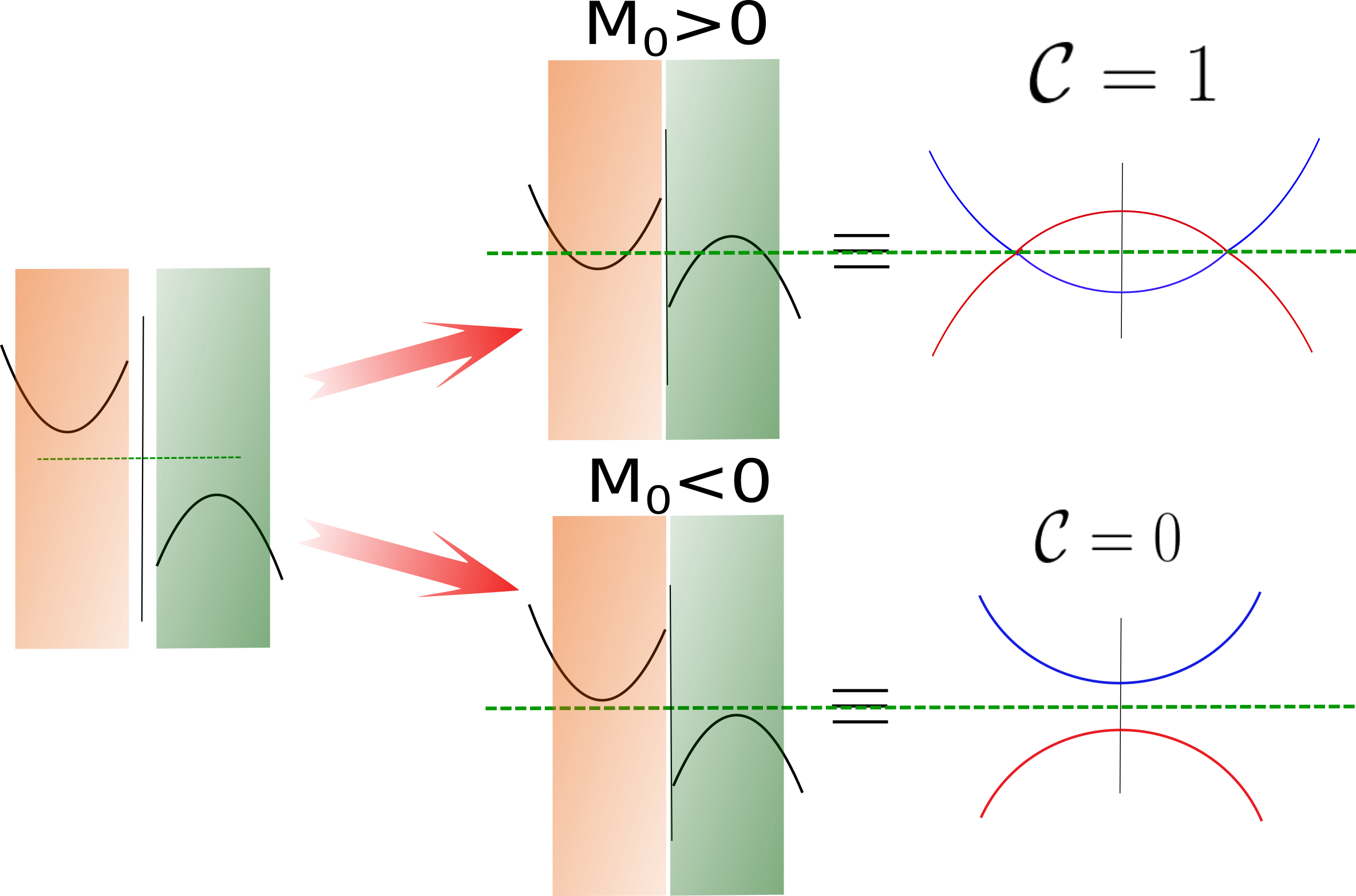}
	\includegraphics[width=42.5mm,height=35.5mm]{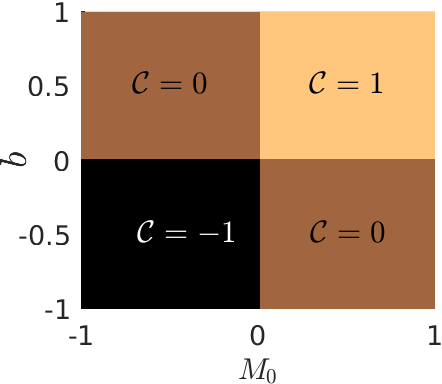}
	\caption{Arising of topological QSH states at the interface of an inverted band InAs/GaSb and HgTe/CdTe quantum well controlling by the band inversion parameter, $M_0$. The figure on the right displays the different QSH topological phases set apart by the topological invariant number $\mathcal{C}=\pm 1$ for conducting QSH states and $\mathcal{C}=0$ for trivial insulating states.}
	\label{Fig1}
\end{figure}

\begin{figure}[t]
	\includegraphics[width=60mm,height=20.5mm]{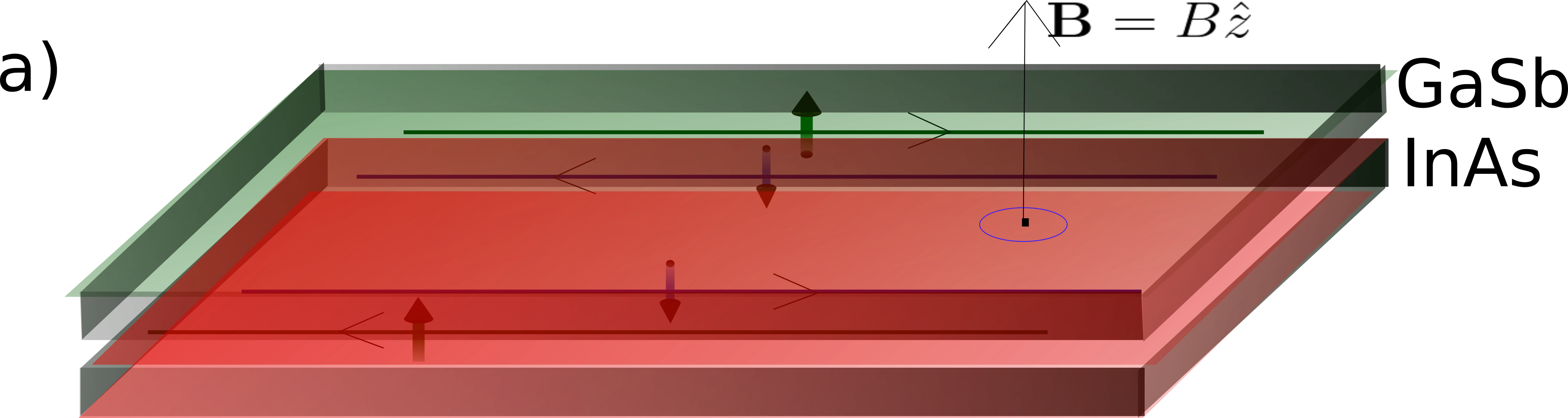}
	\includegraphics[width=80mm,height=55.0mm]{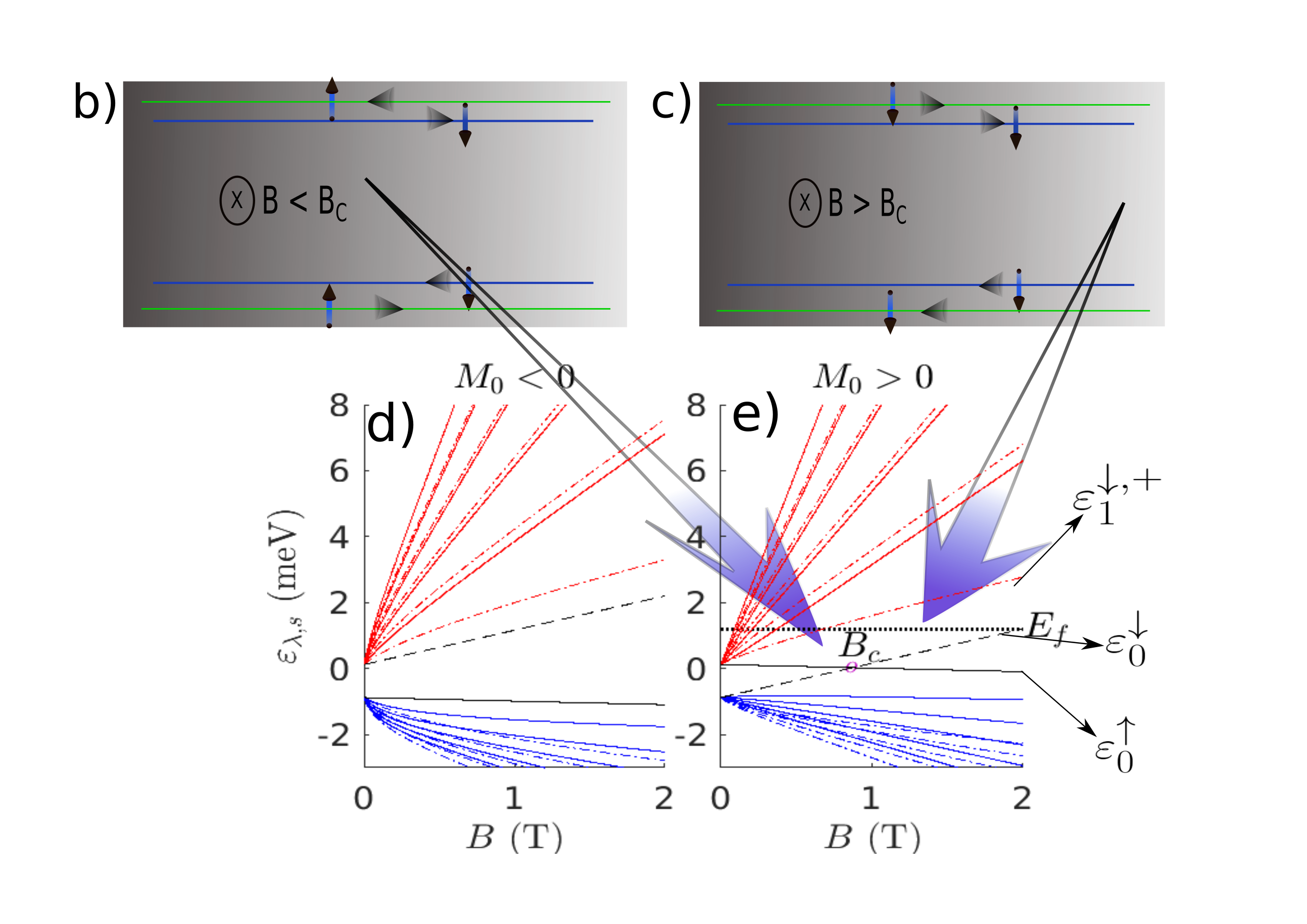}
	\caption{a) A schematic of QSH states in the interface of the inverted quantum well exposing to a magnetic field normal to the 2D plane. b) \& c) Demonstration of the nature of the edge states in the QSH and the QH regime, respectively. d) \& e) Plots of the Landau levels in the entire spectrum when the parameter that controls the band inversion is set as $M_0<0$ [in d)] and $M_0>0$ [ in e)]. The solid and dashed lines are the Landau levels of types $\varepsilon_n^{\downarrow,\lambda}$ and $\varepsilon_n^{\uparrow,\lambda}$, respectively.}
	\label{Fig2}
\end{figure}

Theoretically, the system is described by the Bernevig-Hughes-Zhang (BHZ) Hamiltonian which is a four-band model, having two electron bands (spin up and spin down) and two hole bands (spin up and down).\cite{bhz,Hamil2,TR,Hamil3,ref1,Hamil5,QSH}
\begin{eqnarray}\label{HamilB0}
H=&\epsilon&({\bf k})\sigma_0\tens\sigma_0+M({\bf k})\sigma_0\tens\sigma_0+A k_x\sigma_z\tens\sigma_x\nonumber\\&-&A k_y\sigma_0\tens\sigma_y.
\end{eqnarray}
Here, $\sigma_i$, with $i=x,y,z$ being the Pauli's spin matrices and $\sigma_0$ is the $2\times 2$ identity matrix. The first term $\epsilon({\bf k})=C-dk^2$ consists of the symmetry-breaking term, `$d$' between electron and hole bands. The second term $M({\bf k})=M_0-b k^2$ consists of two parameters, \textit{viz}., `$M_0$' that controls the band inversion and `$b$' that symmetrically controls the band curvatures. The last two terms control the electron-hole coupling. 
This basic theory of the QSH leads to the search of materials with band inversion at the TR invariant k-points, two example of which are InAs/GaSb and HgTe/CdTe quantum wells.
As a consequence of the band inversion, such a system hosts gapless surface or edge states as demonstrated in Fig. (\ref{Fig1}). The spin configuration of the edge states (which always occur in pairs) in such a system is demonstrated in Fig. (\ref{Fig2}) (often called as QSH system). 

It is well-known that an application of a magnetic field is one way of breaking the TR symmetry in such a system\cite{TRB}. In the presence of a magetic field, these inverted quantum systems shows a peculiar behaviour of the zeroth landau levels. 
Precisely, below a critical magnetic field $B<B_c$, the uppermost valence Landau level has instead electron-like characteristics and the lowermost conduction Landau level has hole-like characteristics\cite{ll1,ll2,ll3}. In the regime $B<B_c$ the quantum well still hosts counterpropagating spin polarised edge states. In other words, QSH still exists in the regime $B<B_c$.
Beyond the critical magnetic field ($B>B_c$), there is a band ordering that trace back to the normal Landau levels same as that of a Quantm Hall (QH) system.

In this paper, we present a theoretical study that reveals some results out of the peculiar electron-like nature of the valence band and the hole-like nature of the conduction band. 
Specifically, we present a calculation of the magnetotransport coefficients, namely the longitudinal and Hall conductivity in such a system. Based on the results of the longitudinal and Hall conductivity, we look for some conditions by which one can define a clear distinction between the two possible quantum states in the system, \textit{viz}. , the QSH and QH states. In short, the main finding of this work is centered around the transition from QSH to QH state. We show that, there is a regime of higher Hall conductance around the low Fermi energy level when the system is in the QH regime, while the same is less dominant in the QSH regime. The boundary that partitioned this jump in the conductance, is the critical magnetic field which goes as $B_c=4\hbar M_0/(4b-\mu_B\hbar (g_e+g_h))$. This value of the critical field comprehends the results reported in the previous works \cite{critical1,Hamil2}.

This paper is further organized as follows:  In section \ref{Formalism}, we
present the basic formalism of the inverted band structure in InAs/GaSb and HgTe/CdTe bilayer. In section \ref{Mag_Coef}, we present the analytical calculations for the different transport coefficients.
Numerical results and discussion are presented in section
\ref{RnD}. We summarise this paper in section \ref{Sum}.

\section{Formalism}\label{Formalism}

We started from the BHZ Hamiltonian that describe the QSH states in these systems as given in Eq. (\ref{HamilB0}).
The energy spectrum here is $\varepsilon_\lambda({\bf k})=\epsilon({\bf k})+\lambda\sqrt{M({\bf k})^2+A^2k^2}$, where both the conduction band ($\lambda=+1$) and the valence band ($\lambda=-1$) are spin degenerate. 
The topological invariant number for the Hamiltonian in Eq. (\ref{HamilB0}) is found to be $\mathcal{C}=({\rm sign}(M_0)+{\rm sign}(b))/2$ which does depend on the band inversion parameter, `$M_0$', expectedly.
In the presence of a magnetic field, the momentum vector transformed according to ${\bf k}\rightarrow {\bf k}+e/\hbar {\bf A}$. For an applied magnetic field ${\bf B} = B \hat z$ normal to the plane containing the quantum well, the BHZ Hamiltonian can be expressed in terms of the ladder operator `$a$' and `$a^\dagger$' as below
\begin{eqnarray}
H=\begin{pmatrix}
h_0^+  & -i\frac{\sqrt{2}}{l_c}A a &0 &0\\
i\frac{\sqrt{2}}{l_c}A a^\dagger&h_0^- & 0 & 0\\
0 & 0& h_0^+ &-i\frac{\sqrt{2}}{l_c}A a^\dagger\\
0 & 0& i\frac{\sqrt{2}}{l_c}A a & h_0^-
\end{pmatrix}.
\end{eqnarray}
Here, we have considered the Landau gauge ${\bf{A}}=(0,xB,0)$. The Hamiltonian operator has a two non-zero blocks and it operates seperately on the two spin degrees of freedom. Here,
$h_0^+=-(b+d)\frac{2}{l_c^2}(a^\dagger a+\frac{1}{2})+M_0+C+\frac{g_e\mu_B B}{2}$
and 
$h_0^-= (b-d)\frac{2}{l_c^2}(a^\dagger a+\frac{1}{2})-M_0+C-\frac{g_h\mu_B B}{2}$, with $g_e/g_h$ being the Lande g-factor of the conduction/valence band. The two operators ($a$ and $a^\dagger$) when acting on the oscillator wave function $\vert n\rangle$ give $a\vert n\rangle=\sqrt{n}\vert n-1\rangle$ and $a^\dagger\vert n\rangle=\sqrt{n+1}\vert n+1\rangle$, respectively. The oscillator wavefunction is itself a function of $x-k_yl_c^2$, where $k_y$ is good quantum number, since the Hamiltonian commutes with momentum operator $p_y$. 
The Landau levels wavefunction, $\Psi_{n,k_y}^{s,\lambda}(x,y)$ for $n\ge 1$ in terms of the oscillators wave function $\vert n\rangle$ are as follows
\begin{widetext}
	\begin{eqnarray}\label{wave}
	&&	\Psi_{n,k_y}^{\uparrow, +}(x,y)=\frac{e^{ik_yy}}{\sqrt{\mathcal{A}_{\uparrow, n}}}\begin{pmatrix}
	\vert n-1\rangle\\-i\alpha_{\uparrow n}\vert n\rangle\\0\\0
	\end{pmatrix},
	\Psi_{n,k_y}^{\uparrow,-}(x,y)=\frac{e^{ik_yy}}{\sqrt{\mathcal{A}_{\uparrow, n}}}\begin{pmatrix}
	\alpha_{\uparrow n}\vert n-1\rangle\\i\vert n\rangle\\ 0\\0
	\end{pmatrix},\nonumber\\
	&&
	\Psi_{n,k_y}^{\downarrow,+}(x,y)=\frac{e^{ik_yy}}{\sqrt{\mathcal{A}_{\downarrow, n}}}
	\begin{pmatrix}
	0\\0\\\vert n\rangle\\-i\alpha_{\downarrow n}\vert n-1\rangle
	\end{pmatrix}\text{ and }
	\Psi_{n,k_y}^{\downarrow,-}(x,y)=\frac{e^{ik_yy}}{\sqrt{\mathcal{A}_{\downarrow, n}}}\begin{pmatrix}
	0\\0\\\alpha_{\downarrow n}\vert n\rangle\\i\vert n-1\rangle
	\end{pmatrix}.
	\end{eqnarray}
\end{widetext}
Here $s=+/-$ and $\lambda=+/-$ are the good quantum numbers that denote the spin-up/spin-down electronic or hole state and the conduction/valence bands of the system, respectively.

The corresponding Landau levels are
\begin{eqnarray}
\varepsilon_{n}^{s,\lambda}=&&C+\frac{\mu_B(g_e-g_h)B}{4}+\frac{sb-2dn}{l_c^2}+\lambda\Delta_{s n}.
\end{eqnarray}
with $\mu_Z=\frac{\mu_B(g_e+g_h)B}{4}$ and the different coefficients in Eq. (\ref{wave}) are given by the genral formula
\begin{eqnarray}
\alpha_{s n}=-\frac{\frac{A}{l_c}\sqrt{2n}}{M_0+\frac{sd-2bn}{l_c^2}+\mu_Z+\Delta_{\uparrow n}},
\end{eqnarray}
where $\Delta_{s n}=\sqrt{\frac{2A^2n}{l_c^2}+(M_0+\frac{sd-2bn}{l_c^2}+\mu_Z)^2}$. 
The normalizing constants are
$
\mathcal{A}_{\uparrow,n }=1+\alpha_{\uparrow n}^2,
\; \mathcal{A}_{\downarrow, n}=1+\alpha_{\downarrow n}^2
$.
The eigensystem of the two Landau levels for $n=0$ are found to be as follows
\begin{eqnarray}
{\small
	\left.
	\begin{array}{ll}	
\varepsilon_0^\uparrow=\frac{-d-b}{l_c^2}+M_0+\frac{g_e \mu_B B}{2}+C\\
\varepsilon_0^\downarrow=\frac{-d+b}{l_c^2}-M_0-\frac{g_h \mu_B B}{2}+C
	\end{array}
\right\}}
\end{eqnarray}
\begin{eqnarray}
\Psi_{k_y}^{\uparrow}=\frac{1}{\sqrt{2}}\begin{pmatrix}
\vert 0\rangle\\
0\\
0\\
0
\end{pmatrix}\;\text{ and }
\Psi_{k_y}^{\downarrow}=\frac{1}{\sqrt{2}}\begin{pmatrix}
	0\\
	0\\
	0\\
	\vert 0\rangle
\end{pmatrix}
\end{eqnarray}
The landau level spectrum for both $M_0<0$ and $M_0>0$ is shown in Fig. (\ref{Fig2} [d \& e]). For a given Fermi level (given in solid dashed line of Fig. (\ref{Fig2} [e)]) and assuming that it is located just below the level $\varepsilon_{1}^{\downarrow,+}$, we can divide the spectrum into two regions. Region I, where the QSH states still persist (that the pair of edge states consists of both the up and down spin states: Refer Fig. (\ref{Fig2} [b])). With further increase of the magnetic field, we reach a critical magnetic field $B_c=4\hbar M_0/(4b-\mu_B\hbar (g_e+g_h))$, after which the edge states are just normal QH states (Refer Fig. (\ref{Fig2} [c])). Here the TR symmetry is broken such that the edge states now consist of a pair of electronic states of the same spin configuration propagating in the same direction. We determined this critical field by equating the two Landau level states with $n=0$. For a given values of the involved parameters, this value of the critical magnetic field that separates the QSH and the QH states is found consistent with that reported in references[\cite{critical1,Hamil2}]. 

\section{Derivation of magnetoransport Coefficients}\label{Mag_Coef}
To calculate the longitudinal and Hall components of the conductivity tensor, we employ the Kubo formalism\cite{Mahan}. The collisional contribution accords to the longitudinal component of the conductivity tensor. The diffusive conductivity which has a vanishing contribution is due to the zero value of the expectation value of the diagonal elements of the velocity operator.\cite{kubo}. 

{\bf{Collisional conductivity}}: The calculation of the collisional conductivity involves an assumption that the fermions are elastically scattered by the charged impurities which is presumed to be distributed uniformly over the quantum well. This assumption limits to low temperature for the calculation to remain valid. The Kubo formula for the expression of the collisional conductivity is given by \cite{Van, Carol, vasilo, Peet, wang}
\begin{eqnarray}\label{coll}
	\sigma ^{\rm coll}_{xx}&=&\frac{\beta e^2}{\Omega}\sum\limits_{\xi,\xi^\prime}
	f(\varepsilon_\xi)\{1-f(\varepsilon_{\xi^\prime})\} W_{\xi \xi^\prime}(x^{\xi}-x^{\xi^\prime})^2.
\end{eqnarray}
Here $\vert\xi\rangle=\vert \lambda, s, n,k_y\rangle$ is the set of all quantum numbers that defines the eigenstates of the system, $\Omega$ is the surface area containing the quantum well, $f(\varepsilon_\xi)=1/(\exp((\varepsilon_\xi-\mu)\beta)+1)$ the Fermi-Dirac distribution function with $\beta=1/(k_BT)$ and $x^\xi=\langle \xi\vert x\vert\xi\rangle=k_yl_c^2 $ being the expectation value of the $x$ component of the position operator. Finally, the transition probability between two states $\vert\xi\rangle$ and $\vert\xi^\prime\rangle$ is given by
\begin{eqnarray}\label{trans_prob}
	W_{\xi \xi^{\prime}}&=&\frac{2\pi n_{\rm im}}{\hbar\Omega }
	\sum\limits_{q}\vert U({\bf q})\vert^2\vert F_{\xi,\xi^\prime}\vert^2
	\delta (\varepsilon_{\xi}-\varepsilon_{\xi^\prime}).
\end{eqnarray}
Here the quantity $n_{\rm im}$, is the impurity density in the quantum well and and 	$U(q)=1/(2\epsilon_0 \epsilon(q^2+k^2_s)^{1/2})$ the Fourier transform of the considered screened Yukawa-type impurity potential taken as	$ U(r)=e^2e^{-k_sr}/(4\pi\epsilon_0 \epsilon r)$.  The constants $\epsilon$ and $k_s$ are the dielectric constant of the medium and the screened wave vector, respectively. The form factor $F_{\xi,\xi^\prime}=\langle{\xi}\vert e^{i{\bf q} \cdot{\bf r}}\vert{\xi^\prime}\rangle$ is derived in the Appendix \ref{AppA}.	By writing a general form of the wavefunction as $\Psi_{n,k_y}^{s,\lambda}(x,y)=e^{ik_yy}\begin{pmatrix}
p_n\vert n-1\rangle&iq_n\vert n\rangle& r_n\vert n-1\rangle&is_n\vert n\rangle
\end{pmatrix}^\prime/\sqrt{\mathcal{A}_{s,n}}$, where `$\prime$' indicates the transpose, the collisional conductivity for $n\ge 1$ as given in Eq. (\ref{A6}) of the appendix is
\begin{eqnarray}\label{coll1}
	\frac{\sigma_{xx}^{\rm coll}}{\sigma_Q}&&=\frac{\beta \Gamma}{4\pi}\sum_{n,s,\lambda}\frac{f(\varepsilon_n^{s,\lambda})[1+f(\varepsilon_n^{s,\lambda})]}{\mathcal{A}_{s,n}^2}\{(1+2n)(q_n^2+r_n^2)^2\nonumber\\
	&&-2ns_n^2(q_n^2+r_n^2)-2p_n^2[s_n^2+n(q_n^2+r_n^2-2s_n^2)]\nonumber\\&&+(2n-1)(p_n^4+s_n^4)\}.
\end{eqnarray}
Here in this paper, we will scale the longitudinal and Hall conductivity in units of $\sigma_Q=e^2/h$. Following Eq. (\ref{n_0_coll}) of the Appendix (\ref{AppA}), we have for $n=0$
\begin{eqnarray}\label{coll2}
\frac{\sigma_{xx}^{\rm coll}}{\sigma_Q}=\frac{\beta\Gamma}{4}[f(\varepsilon_0^\uparrow)\{1+f(\varepsilon_0^\uparrow)\}+f(\varepsilon_0^\downarrow)\{1+f(\varepsilon_0^\downarrow)\}].
\end{eqnarray}

{\bf Hall conductivity:}
The expression for the Hall conductivity  $\sigma_{yx}$ is given by \cite{vasilo, Peet, wang}
\begin{eqnarray}\label{hall_con}
	\sigma_{yx} & = & \frac{i\hbar e^2}{\Omega }\sum \limits_{\xi \neq \xi^{\prime}}
	\langle \xi\vert v_y\vert\xi^{\prime} \rangle
	\langle \xi^{\prime}\vert v_x\vert\xi \rangle\frac{f(\varepsilon-_{\xi}) - 
		f(\varepsilon_{\xi^{\prime}})}{(\varepsilon_{\xi}-\varepsilon_{\xi^\prime})^2}.
\end{eqnarray}
The matrix elements of the components of the velocity operator are given 
in Eqn (\ref{vell}) of Appendix \ref{AppB}. By virtue of the Kronecker delta symbols in 
Eqs. (\ref{vy_exp} \& \ref{vx_exp}), it is confirmed that the 
transitions are allowed only between the adjacent Landau 
levels $n^\prime = n \pm 1 $. The contribution due to the inter-spin branch transitions is found to be negligibly small compared to the contribution from the intra-spin branch transitions. Also, we observed that the contribution of the transitions from conduction to valence bands and vice versa of the type $\varepsilon_{n}^{s,\pm}\rightarrow \varepsilon_{n+1}^{s,\mp}$ are again found to be negligibly small. The  Hall conductivity in Eq. (\ref{hall_con}) thus has terms that should only involve transitions of the type $\varepsilon_{n}^{s,\pm}\rightarrow \varepsilon_{n+1}^{s,\pm}$. 

The form of the Hall conductivity expressed as a summation involving the quantum numbers ($n,s,\lambda$) is derived in Eq. (\ref{hall_con_form_1} \& \ref{hall_con_form_2}) of Appendix [\ref{AppB}], the results of which will be discussed in the subsequent section (\ref{RnD}).
\begin{figure}[b]
	\includegraphics[width=86mm,height=50.5mm]{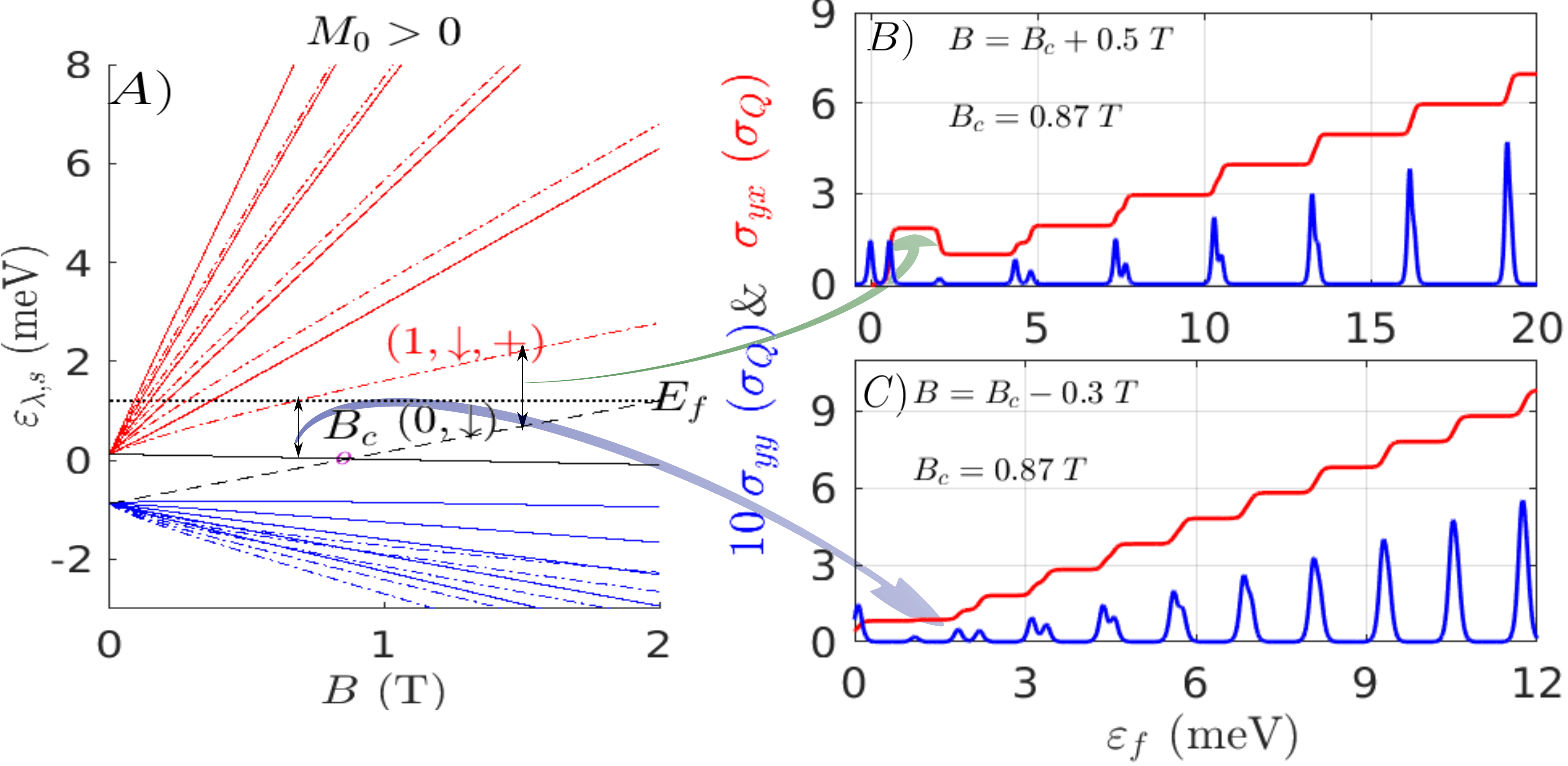}
	\caption{A). Plots of the Landau levels as a function of the magnetic field. B) \& C). Plots of the longitudinal and Hall conductivity as a function of the Fermi energy for cases where $B=B_C+0.5T$ and $B=B_C-0.3T$, respectively.}
	\label{Fig3}
\end{figure}

\begin{figure}[http!]
	\includegraphics[width=44.5mm,height=40.5mm]{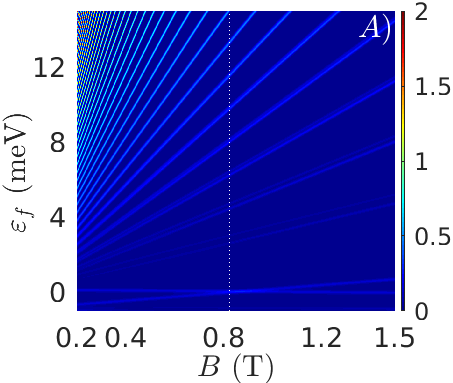}
	\includegraphics[width=41.0mm,height=40.0mm]{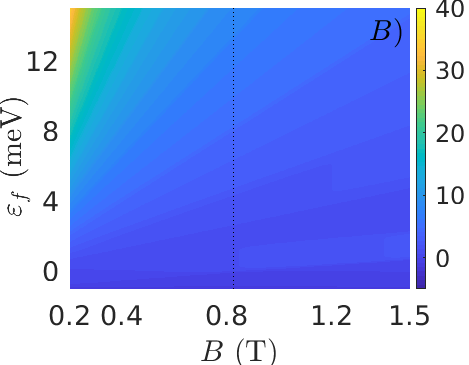} 
	\caption{A) \& B) Plots of the longitudinal and Hall conductivity (in units of $e^2/h$) as a function of the Fermi energy and the magnetic field, respectively. The vertical dotted line shows the critical field which separates the two regime describing the QSH and QH states.}
	\label{Fig4}
\end{figure}

\section{Results and Discussions}\label{RnD}
For analysing the Longitudinal and Hall conductivity in the above section \ref{Mag_Coef}, we take the value of the different parameters refering to table I below. 
\begin{table}[http!]
	\begin{tabular}{|l|l|l|l|l|l|}
		\hline
		\thead{$b$\\ (eV nm$^2$)}  & 	\thead{$d$ \\(eV nm$^2$)}& \thead{$C$ \\(eV)}& \thead{$A$ \\(eV nm)}& \thead{$M_0$ \\(meV)}\\ \hline
		$\thead{40}$  & $\thead{-30}$ & $\thead{-0.375\times 10^{-4}}$ & \thead{$0.03$}   & \thead{$0.5$}\\ \hline
	\end{tabular}
	\label{t11}
	\caption{The different parameters associated with the BHZ Hamiltonian in eq. (\ref{HamilB0})}.
\end{table}
We restricted our analysis to the regime where $M_0>0$, such that the system would always host the zeroth Landau level crossing as shown in Fig. (\ref{Fig3} [A]).
The distinction between the different quantum states was previously done by inspecting the additional absorption peak at low energies in the QSH regime and that the peak vanishes in the QH regime if the Fermi level is situated in the bulk gap\cite{critical1}. In this paper, we show this kind of partition on the basis of the different magnetotransport parameters, namely, the longitudinal and Hall conductivity.

As indicated earlier, the clear distinction between the QSH and QH states is the critical field, $B_c=4\hbar M_0/(4b-\mu_B\hbar (g_e+g_h))$. To reckon with this presupposed conclusion, we plot both the longitudinal and Hall conductivity in Fig. (\ref{Fig3} [B \& C]) as a function of the Fermi energy at two values of the magnetic field around the critical field (precisely at $B=B_c+0.2$  T Fig. (\ref{Fig3} [B]) and $B=B_c+0.2$ T Fig. (\ref{Fig3} [C])). A closed observation of the longitudical conductivity shows an expected peak around $\varepsilon_f=\varepsilon_0^\uparrow(B_c)=\varepsilon_0^\downarrow(B_c)$. The Hall conductivity shown in the same figure carries more information about the quantum state of the system. Its closed observation around $\varepsilon_f=\varepsilon_0^\uparrow(B_c)=\varepsilon_0^\downarrow(B_c)$ for $B>B_c$ (the QH regime) shows that the component of the conductivity responds by showing a larger magnitute in comparison to that for $B<B_c$ (the QSH regime). The little bump of the Hall conductivity comes from those type of transition $\varepsilon_0^\downarrow\rightarrow \varepsilon_1^{\downarrow,-}$ which clearly dominates the spectrum in the QH regime.

The above observation prompts us to look into the behaviour of the above magnetotransport parameters as a function of the magnetic field and the Fermi energy. We show the 2D plots that address this notion in Fig. (\ref{Fig4}). Figure (\ref{Fig4} [A]) shows the longitudinal conductivity as a function of the magnetic field and Fermi energy which expectedly follows the Landau level spectrum with a very sharp peak when the Fermi energy is same as the zeroth Landau level at $B=B_c$. As indicated above, the Hall conductivity carries information about the nature of the topological state of the system. The Hall conductivity in Figure (\ref{Fig4} [B]) bumps up in the QH regime that begins from a magnetic field, $B=B_c$. This critical magnetic field is thus a boundary below which the system would still host QSH edge states and beyond which the edge states revert to the normal QH states. Furthermore, the Hall plateaus show their robustness for Fermi energy situated in the bulk gap of the inverted quantum well which can be seen in the regime of higher Fermi energy. 

\section{Conclusion}\label{Sum}
We have presented a theoretical study of the magnetotransport properties of an inverted band in the quantum well of InAs/GaSb and HgTe/CdTe heterostructure. The inverted band in these heterostructures is described by the Bernevig-Hughes-Zhang Hamiltonian that perfectly described the quantum spin Hall (QSH) states discovered in such a system. In presence of a magnetic field, there are two possible quantum states, namely, the QSH and Quantum Hall (QH) states. We found the value of the critical field $B_c=4\hbar M_0/(4b-\mu_B\hbar (g_e+g_h))$ that partitioned the spectrum into these two types of quantum states. The paper is centered around the value of this critical field where the spectrum shows electron-like behavior of the hole states and hole-like behavior of the electron states in the QSH regime. The robustness of the QH states in these system is evident when the Fermi energy overshoot any energy level between $\varepsilon_0^\downarrow$ and $ \varepsilon_1^{\downarrow,-}$. When the Fermi level falls in between the Landau levels $\varepsilon_0^\downarrow$ and $ \varepsilon_1^{\downarrow,-}$, the Hall conductivity bumps up in the QH regime.
In a nutshell, the Hall conductivity provides a signature to identify the QSH and QH regime as one varies the gate voltage in an inverted quantum well.

\textit{Acknowledgments}: This work is an outcome of
the Research work carried out under the DST-INSPIRE project DST/INSPIRE/04/2019/000642, Government of India.

\appendix
\section{Derivation of the collisional conductivity}\label{AppA}
Since the term $F_{\xi,\xi^\prime}$ is proportional to 
$\delta_{k_y^\prime,k_y+q_y}$, the summation over 
$k_y^\prime$ in Eq. (\ref{coll}) can be easily evaluated with the 
replacement of $k_y^\prime$ by $k_y+q_y$ and we have 
$(x^\xi-x^{\xi^\prime})=q_y^2l_c^4=q^2l_c^4\sin^2\phi$. 
The delta function in Eq. (\ref{trans_prob}), 
$\delta (\varepsilon_{\xi}-\varepsilon_{\xi^\prime}) = 
\delta (\varepsilon_n^{s\lambda}-\varepsilon_{n^\prime}^{s^\prime\lambda^\prime})$, 
ensures the possibilities of only intra-branch and 
intra-level scattering i.e. $n^\prime=n$, $s^\prime=s$ and $\lambda^\prime=\lambda$. 
Again $\delta (E_n^\lambda-E_{n^\prime}^{\lambda^\prime})$ can be written
in its usual Lorentzian representation i.e. 
$\delta (E_n^\lambda-E_{n^\prime}^{\lambda^\prime})
=(1/\pi)\Gamma/[(E_n^\lambda-E_{n^\prime}^{\lambda^\prime})^2 + 
\Gamma^2]$  with $\Gamma$ is the impurity induced Landau level broadening. 
We also have $\sum_{k_y}\rightarrow \Omega/(2\pi l_c^2)$ and
$\sum_{{\bf q}}\rightarrow (\Omega/(2\pi)^2)\int qdqd\phi$. So by 
inserting Eq. (\ref{trans_prob}) into Eq. (\ref{coll}) and after doing all 
the summations one can obtain the following expression of the collisional 
conductivity:

\begin{eqnarray}\label{coll1_1}
\sigma_{xx}^{\rm coll}&=&\frac{e^2}{h}\frac{\beta n_il_c^2U_0^2}{2\pi\Gamma}
\sum_{n,\lambda}f(\varepsilon_n^\lambda)\{1-f(\varepsilon_n^\lambda)\}\nonumber\\
&\times&\int dqq^3\vert F_{nn}^\lambda(q)\vert^2.
\end{eqnarray}
In deriving Eq. (\ref{coll1_1}) we have used the following approximation
$\vert U(q) \vert \simeq U_0=e^2/(2\epsilon_0\epsilon k_s)$ since $q\ll k_s$. Now 
using the fact $n_{\rm im} U_0^2 \sim (\Gamma l_c)^2/(4\pi)$, we finally have
\begin{eqnarray}\label{A2}
\frac{\sigma_{xx}^{\rm coll}}{\sigma_Q}=\frac{\beta \Gamma}{4\pi}
\sum_{n,s,\lambda}f(\varepsilon_n^{s,\lambda})\{1-f(\varepsilon_n^{s,\lambda})\}I_n^{s,\lambda},
\end{eqnarray}
where
\begin{eqnarray}\label{A3}
I_n^{s,\lambda}&=&l_c^4\int_0^\infty q^3dq\vert\mathcal{F}_n^{s\lambda}\vert^2
\end{eqnarray}
The form factor, $\mathcal{F}_n^{s\lambda}$ is given by $\mathcal{F}_n^{s\lambda}=\langle{\Psi_n^{s,\lambda}}^\dagger(x,y)\vert e^{i{\bf q}\cdot{\bf r}}\vert{\Psi_n^{s,\lambda}}(x,y)\rangle$. For derivation of the form factor, we write the wavefunction as
\begin{eqnarray}
\Psi_n^{s,\lambda}(x,y)=\frac{e^{ik_yy}}{\sqrt{\mathcal{A}_{s,n}}}\begin{pmatrix}
p_n\vert n-1\rangle\\
iq_n \vert n\rangle\\
r_n \vert n\rangle\\
i s_n\vert n-1\rangle 
\end{pmatrix},
\end{eqnarray}
which eventually gives us
\begin{eqnarray}
\vert \mathcal{F}_n^{s\lambda}\vert^2&&=\frac{e^{-u}}{{\mathcal{A}_{s,n}}^2}[p_n^2\mathcal{L}_{n-1}(u)+q_n^2\mathcal{L}_n(u)\nonumber\\&&+r_n^2\mathcal{L}_n(u)+s_n^2\mathcal{L}_{n-1}(u)]^2
\end{eqnarray}
where $u=q^2l_c^2/2$.
The integral in Eq. (\ref{A3}) is now
\begin{eqnarray}
I_n^{s,\lambda}&=&\int_0^\infty 2udu
\frac{e^{-u}}{{\mathcal{A}_{s,n}}^2}[p_n^2\mathcal{L}_{n-1}(u)+q_n^2\mathcal{L}_n(u)\nonumber\\&&+r_n^2\mathcal{L}_n(u)+s_n^2\mathcal{L}_{n-1}(u)]^2
\end{eqnarray}
Using the recursive relation
$
u\mathcal{L}_n(u)=(2n+1)\mathcal{L}_n(u)-n\mathcal{L}_{n-1}(u)-(n+1)\mathcal{L}_{n+1}(u)
$, one can arrive at 
\begin{eqnarray}
&&I_n^{s,\lambda}=\frac{1}{{\mathcal{A}_{s,n}}^2}\Big[
(1+2n)(q_n^2+r_n^2)^2-2ns_n^2(q_n^2+r_n^2)\nonumber\\&&+(2n-1)(p_n^4+s_n^4)-2p_n^2[s_n^2+n(q_n^2+r_n^2-2s_n^2)]\Big]
\end{eqnarray}
thus giving us the form of the longitudinal conductivity to be
\begin{eqnarray}\label{A6}
\frac{\sigma_{xx}^{\rm coll}}{\sigma_Q}&&=\frac{\beta \Gamma}{4\pi}\sum_{n,s,\lambda}\frac{1}{{\mathcal{A}_{s,n}}^2}f(\varepsilon_n^{s,\lambda})[1+f(\varepsilon_n^{s,\lambda})]\{-2ns_n^2(q_n^2+r_n^2)\nonumber\\
&&+(2n-1)(p_n^4+s_n^4)-2p_n^2[s_n^2+n(q_n^2+r_n^2-2s_n^2)]\nonumber\\&&+(1+2n)(q_n^2+r_n^2)^2\}.
\end{eqnarray}
In the above equation, we have made the substitution $n_{\rm imp}U_0^2=(\Gamma l_c^2)/4\pi$.
For the $n=0$ Landau states, we have 
\begin{eqnarray}\label{n_0_coll}
\frac{\sigma_{xx}}{\sigma_Q}=\frac{\beta\Gamma}{4}\sum_{s}f(\varepsilon_0^s)[1+f(\varepsilon_0^s)].
\end{eqnarray}

\section{Derivation of the Hall Conductivity}\label{AppB}
Starting from Eq. \ref{hall_con}, we will continue with calculation of the diagonal components of the velocity matrix elements. Using Heisenberg equation of motion $v_i = (1/i\hbar)[x_i,H]$, we calculate the
following components of the velocity operators
\begin{eqnarray}\label{vell}
\left.
\begin{array}{ll}
v_x=\frac{1}{\hbar}\begin{pmatrix}
V_x^1 & \mathcal{O}_{2\times 2}\\
\mathcal{O}_{2\times 2} &V_x^2
\end{pmatrix}
\\
v_y=\frac{1}{\hbar}\begin{pmatrix}
V_y & \mathcal{O}_{2\times 2}\\
\mathcal{O}_{2\times 2} &V_y
\end{pmatrix}
\end{array}
\right\}
\end{eqnarray}
where 
\begin{eqnarray*}
V_y=\frac{1}{\hbar}\begin{pmatrix}
\sqrt{2}\frac{b+d}{l_c^2}(a^\dagger+a) & i A\\
-i A & -\sqrt{2}\frac{b-d}{l_c^2}(a^\dagger+a)
\end{pmatrix}\\
V_x^1=
\begin{pmatrix}
	-i\sqrt{2}\frac{b+d}{l_c}(a^\dagger-a) & A\\
	A & i\sqrt{2}\frac{b-d}{l_c}(a^\dagger-a)
\end{pmatrix}\\
V_x^2=
\begin{pmatrix}
		-i\sqrt{2}\frac{b+d}{l_c}(a^\dagger-a) & -A\\
		-A & i\sqrt{2}\frac{b-d}{l_c}(a^\dagger-a)
	\end{pmatrix}
\end{eqnarray*}
and $\mathcal{O}_{2\times 2}$ is a $2\times 2$ null matrix.

The quantities $\langle\Psi_{m,k_y}^{s^\prime,\lambda^\prime}\vert v_y\vert \Psi_{n,k_y}^{s,\lambda}\rangle$ and $\langle\Psi_{m,k_y}^{s^\prime,\lambda^\prime}\vert v_x\vert \Psi_{n,k_y}^{s,\lambda}\rangle$ now become
\begin{eqnarray}\label{vy_exp}
&&\langle\Psi_{n,k_y}^{s,\lambda}\vert v_y\vert \Psi_{m,k_y}^{s^\prime,\lambda^\prime} \rangle=\frac{1}{\hbar\sqrt{\mathcal{A}_m^{s^\prime,\lambda^\prime}\mathcal{A}_n^{s,\lambda}}}\Big[\Big(\sqrt{2n}\frac{b+d}{lc}p_np_m\nonumber\\
&&-\sqrt{2n}\frac{b-d}{l_c}s_ns_m-\sqrt{2(n+1)}\frac{b-d}{l_c}q_nq_m\nonumber\\&&+\sqrt{2(n+1)}\frac{b+d}{l_c}r_nr_m-Aq_np_m-Ar_ns_m\Big)\delta_{m,n+1}\nonumber\\+
&&\Big(\sqrt{2(n-1)}\frac{b+d}{lc}p_np_m-\sqrt{2(n-1)}\frac{b-d}{l_c}s_ns_m\nonumber\\&&
-\sqrt{2n}\frac{b-d}{l_c}q_nq_m+\sqrt{2n}\frac{b+d}{l_c}r_nr_m\nonumber\\&&-Aq_mp_n-Ar_ms_n\Big)\delta_{m,n-1}
\Big]
\end{eqnarray}
and
\begin{eqnarray}\label{vx_exp}
&&\langle\Psi_{m,k_y}^{s^\prime,\lambda^\prime}\vert v_x\vert \Psi_{n,k_y}^{s,\lambda}\rangle=\frac{i}{\hbar\sqrt{\mathcal{A}_m^{s^\prime,\lambda^\prime}\mathcal{A}_n^{s,\lambda}}}\Big[\Big(\sqrt{2n}\frac{b+d}{lc}p_np_m\nonumber\\
&&-\sqrt{2n}\frac{b-d}{l_c}s_ns_m-\sqrt{2(n+1)}\frac{b-d}{l_c}q_nq_m\nonumber\\&&+\sqrt{2(n+1)}\frac{b+d}{l_c}r_nr_m-Aq_np_m-Ar_ns_m\Big)\delta_{m,n+1}\nonumber\\-
&&\Big(\sqrt{2(n-1)}\frac{b+d}{lc}p_np_m-\sqrt{2(n-1)}\frac{b-d}{l_c}s_ns_m\nonumber\\&&
-\sqrt{2n}\frac{b-d}{l_c}q_nq_m+\sqrt{2n}\frac{b+d}{l_c}r_nr_m\nonumber\\&&-Aq_mp_n-Ar_ms_n\Big)\delta_{m,n-1}
\Big]
\end{eqnarray}
This leads to the expression of the Hall conductivity to be given as under,
\begin{eqnarray}\label{hall_con_form_1}
&&\frac{\sigma_{yx}}{\sigma_Q}=\sum_{n,s,\lambda}\frac{2}{\mathcal{A}_{s,n}\mathcal{A}_{s,n+1}}\frac{f(\varepsilon_n^{s,\lambda})-f(\varepsilon_{n+1}^{s,\lambda})}{(\varepsilon_{n}^{s,\lambda}-\varepsilon_{n+1}^{s,\lambda})^2}\nonumber\\
&&\Big(\sqrt{2n}\frac{b+d}{lc}p_np_{n+1}
-\sqrt{2n}\frac{b-d}{l_c^2}s_ns_{n+1}
\nonumber\\&&
-\sqrt{2(n+1)}\frac{b-d}{l_c^2}q_nq_{n+1}+\sqrt{2(n+1)}\frac{b+d}{l_c^2}r_nr_{n+1}\nonumber\\&&-\frac{A}{l_c}q_np_{n+1}-\frac{A}{l_c}r_ns_{n+1}\Big)^2
\end{eqnarray}

For the contribution coming from the transitions $(0,\downarrow)\rightarrow (1,\uparrow,+)$, $(0,\downarrow)\rightarrow (1,\downarrow,+)$, $(0,\uparrow)\rightarrow (1,\uparrow,+)$ and $(0,\uparrow)\rightarrow (1,\downarrow,+)$, the Hall conductivity is given by
\begin{eqnarray}\label{hall_con_form_2}
\left.
\begin{array}{ll}
\frac{\sigma_{yx}}{\sigma_Q}=\sum_s\frac{1}{\mathcal{A}_{s,1}}\Big[\frac{A }{l_c}p_1+\sqrt{2}\frac{b-d}{l_c^2}q_1\Big]^2\frac{f(\varepsilon_{0}^{\downarrow})-f(\varepsilon_{1}^{s,\lambda})}{(\varepsilon_{1}^{s,\lambda}-\varepsilon_{0}^{\downarrow})^2}\\
\frac{\sigma_{yx}}{\sigma_Q}=\sum_s\frac{1}{\mathcal{A}_{s,1}}\Big[\frac{A }{l_c}s_1-\sqrt{2}\frac{b-d}{l_c^2}r_1\Big]^2\frac{f(\varepsilon_{0}^{\uparrow})-f(\varepsilon_{1}^{s,\lambda})}{(\varepsilon_{1}^{s,\lambda}-\varepsilon_{0}^{\uparrow})^2}
\end{array}
\right\}
\end{eqnarray}

\end{document}